\documentclass[11pt]{article}
\usepackage{moriond,epsfig}

\def\ra{\rightarrow}

\def\be{\begin{equation}}
\def\ee{\end{equation}}
\def\bea{\begin{eqnarray}}
\def\eea{\end{eqnarray}}

\begin{document}
\vspace*{4cm}
\title{Top-quark pair cross-section measurement in the lepton+jets channel}

\author{ M. Pinamonti\\
on behalf of the ATLAS collaboration}

\address{INFN Udine \& University of Trieste,\\
Strada Costiera 11, 34151 Trieste, Italy}

\maketitle
\abstracts{
A measurement of the production cross-section for top quark pairs in $pp$ collisions at $\sqrt{s} = $7~TeV is presented
using data recorded with the ATLAS detector at the Large Hadron Collider (LHC).
Events are selected in the lepton+jets topology by requiring a single lepton (electron or muon),
large missing transverse energy and at least three jets.
No explicit identification of secondary vertices inside jets ($b$-tagging) is performed.
In a data sample of 35.3 pb$^{-1}$,
2009 and 1181 candidate events are observed in the $\mu$+jets and $e$+jets topology, respectively.
A simple multivariate method using three kinematic variables is employed 
to extract a cross-section measurement of 171$\pm$17(stat.)$^{+20}_{-17}$(syst.)$\pm$5(lumi.) pb.
}

\section{Introduction}

A precise measurement of the top-pair ($t\bar{t}$) inclusive cross-section at this early stage of the LHC data taking 
is of central importance for several reasons.

First of all it allows a direct comparison with theoretical calculations 
providing a precision test of the predictions of perturbative QCD. 
Additionally $t\bar{t}$ production is an important background in many searches for physics beyond the Standard Model,
and new physics may also give rise to additional $t\bar{t}$ production mechanisms or modifications of
the top quark decay channels.
Finally, this is one of the first precision measurements 
implying the reconstruction of final states 
including jets, electrons ($e$), muons ($\mu$) and missing transverse energy 
($E_T^{miss}$), 
and since many models of physics beyond the 
Standard Model predict events with similar signatures 
it provides an essential stepping stone toward the identification of new physics. 

\section{Top-pair production and decay}

In the Standard Model (SM) the $t\bar{t}$ production cross-section in $pp$ collisions is calculated to be
165$^{+11}_{-16}$ pb at a centre of mass energy of $\sqrt{s} = $7 TeV assuming a top mass of 172.5 GeV \cite{xsec}. 
Top quarks are predicted to decay into a $W$ boson and a $b$-quark ($t \ra Wb$) nearly 100$\%$ of the time.
Depending on the decays of the two $W$ bosons into a pair of quarks ($W \ra q\bar{q}'$) 
or a lepton-neutrino pair ($W \ra \ell\nu$), events with a $t\bar{t}$ pair can be classified as:
\begin{itemize}
 \item dilepton: when both $W$s decay leptonically; 
 \item single-lepton: when one of the $W$ decays leptonically and the second one hadronically;
 \item all-hadronic: when both $W$s decay into quarks.
\end{itemize}

For the analysis reported here single-lepton $t\bar{t}$ events are selected, 
considering only events with exactly one electron ($e$+jets channel) 
and exactly one muon ($\mu$+jets channel) 
and without using any $b$-tagging information. 
A more detailed description of the analysis can be found in \cite{ljets_pretag}.

Other complementary analyses are performed in ATLAS to extract the $t\bar{t}$ production cross-section in dilepton \cite{dilep}
and all-hadronic \cite{allhad} channels
as well as in the single-lepton channel making use of the $b$-tagging information \cite{ljets_btag}.

\section{Event Selection}

To select $t\bar{t}$ events in the single lepton final state, 
the following event selections are applied:
\begin{itemize}
 \item the appropriate single-electron or single-muon trigger has fired;
 \item the event contains exactly one reconstructed lepton (electron or muon) with $p_T > $20 GeV,
matching the corresponding high-level trigger object;
 \item if a muon is reconstructed, $E_T^{miss} > $20 GeV and $E_T^{miss} + m_T(W) > $60 GeV is required
\footnote{
Here $m_T(W)$ is the $W$-boson transverse mass, defined as $\sqrt{2p_T^{\ell}p_T^{\nu}(1-cos(\phi_\ell - \phi_\nu))}$
where the measured missing $E_T$ vector provides the neutrino information.};
 \item if an electron is reconstructed, 
$E_T^{miss} > $35 GeV and $mT (W) > $25 GeV are required;
 \item the event is required to have $\geq$ 3 jets with $p_T > $25 GeV and $|\eta| < $2.5.
\end{itemize}
Depending on the flavour of the lepton ($e$ or $\mu$) and on the number of reconstructed jets (exactly three or at least four) 
the events are classified as $e$+3-jets, $\mu$+3-jets, $e$+$\geq$4-jets or $\mu$+$\geq$4-jets, 
giving rise to four statistically independent channels.

\section{Background treatment}

The most important backgrounds 
after the event selections described above
are:
\begin{itemize}
 \item the production of a $W$ boson in association with jets ($W$+jets),
 \item the production of QCD multi-jet events 
in which a fake or non-prompt lepton is reconstructed as a real prompt electron or muon,
 \item other minor backgrounds including single top electro-weak production, 
$Z$+jets and diboson ($WW$,$WZ$ and $ZZ$) events.
\end{itemize}

The number of events observed in data and predicted by simulation or by data-driven estimates 
in each of the four channels are given in Table \ref{tab:events}.

\begin{table}

\begin{minipage}[t]{0.47\textwidth}
 
\begin{scriptsize}  \centering
  \begin{tabular}{|c|cc|cc|}
\hline
events	&$e$ &$e$ &$\mu$ &$\mu$ \\
 & +3-jets	&+$\geq$4-jets	&+3-jets	&+$\geq$4-jets	\\
\hline\hline
 $t\bar{t}$	& 116	& 194	& 161	& 273 \\
 QCD & 62 & 22 & 120 & 51 \\
 $W$+jets & 580 & 180 & 1100 & 310 \\
 $Z$+jets	& 32	& 18	& 70	& 25 \\
 single $t$	& 22	& 11	& 32	& 15 \\
 diboson	& 9	& 3	& 16	& 4 \\
\hline\hline
 Data		& 781	& 400	& 1356	& 653 \\
\hline
 \end{tabular}
\end{scriptsize}
 \caption{Numbers of events in the four selection channels. 
The observed data events are shown, together with the MC simulation estimates for $t\bar{t}$,
$W$+jets, $Z$+jets and single-top and diboson events, normalised to the data integrated luminosity of 35 pb$^{-1}$. 
The data-driven estimates for QCD multi-jet background are also shown.\hfill}
 \label{tab:events}

\end{minipage}
\hfill
\begin{minipage}[t]{0.47\textwidth}

\begin{scriptsize}\centering
 \begin{tabular}{|c|c|}
\hline
  Source	& $\Delta\sigma/\sigma$[\%]	\\
  \hline\hline
  Statistical uncertainty & $\pm$9.7 \\
  \hline\hline
  $\ell$ reco., id. and trigger & -1.9 / +2.6 \\
  Jet energy scale, resolution and reco. & -6.1 / +5.7 \\
  QCD normalization & $\pm$3.9 \\
  QCD shape & $\pm$3.4 \\
  $W$+jets shape & $\pm$1.2 \\
  Other backgrounds & $\pm$0.5 \\
  ISR/FSR & -2.1 / +6.1 \\
  PDFs & -3.0 / +2.8 \\
  Parton shower generator & $\pm$3.3 \\
  Monte Carlo generator & $\pm$2.1 \\
  Limited MC statistics & $\pm$1.8 \\
  Pile-up & $\pm$1.2 \\
  \hline
  Total systematics & -10.2 / +11.6 \\
  \hline\hline
  Luminosity & $\pm$3.4 \\
\hline
 \end{tabular}
\end{scriptsize}

\caption{List of the main sources of uncertainty affecting the final measurement. 
For each of the listed sources the relative effect on the measured $t\bar{t}$ cross-section 
expressed as relative uncertainty is reported.\hfill}
\label{tab:syst}

\end{minipage}

\end{table}

The different backgrounds are treated in different ways to determine the shape and the normalization 
of the kinematical distributions used to build the likelihood discriminant to extract the cross-section measurement.
For the $W$+jets background the shapes are taken from Monte Carlo (MC) simulation, 
while the normalization is extracted from the fit 
(see Section 5). 
For the QCD multi-jet background both the shapes and the normalization are extracted 
with data-driven methods. 
For the other backgrounds, both the shapes and the normalization are taken from MC simulation.

\section{Cross-Section Measurement}

The $t\bar{t}$ production cross-section is extracted by exploiting the different properties of $t\bar{t}$ events with
respect to the dominant $W$+jets background. 
Three variables were selected 
for their discriminant power, for the small correlation between them 
and by considering the effect of the jet energy scale uncertainty.

These variables are:
\begin{itemize}
 \item the pseudorapidity of the lepton $\eta_{lepton}$, which exploits the fact that $t\bar{t}$ events produce more central
leptons than $W$+jet events;
 \item the charge of the lepton $q_{lepton}$, which uses the fact that 
$t\bar{t}$ events produce charge-symmetric leptons
while $W$+jet events produce an excess of positively charged leptons;
 \item the exponential of the aplanarity ($exp(−8\times A)$), 
\footnote{Here $A = \frac{3}{2} \lambda_3$, where $\lambda_3$ is the smallest
eigenvalue of the normalized momentum tensor calculated using the selected jets and lepton in the
event.}
which exploits the fact that $t\bar{t}$ events are more isotropic than $W$+jets.
\end{itemize}

\begin{figure}[t]

\begin{minipage}[t]{0.5\textwidth}
\centering
(a)\\
\includegraphics[width=0.9\textwidth,keepaspectratio=true]{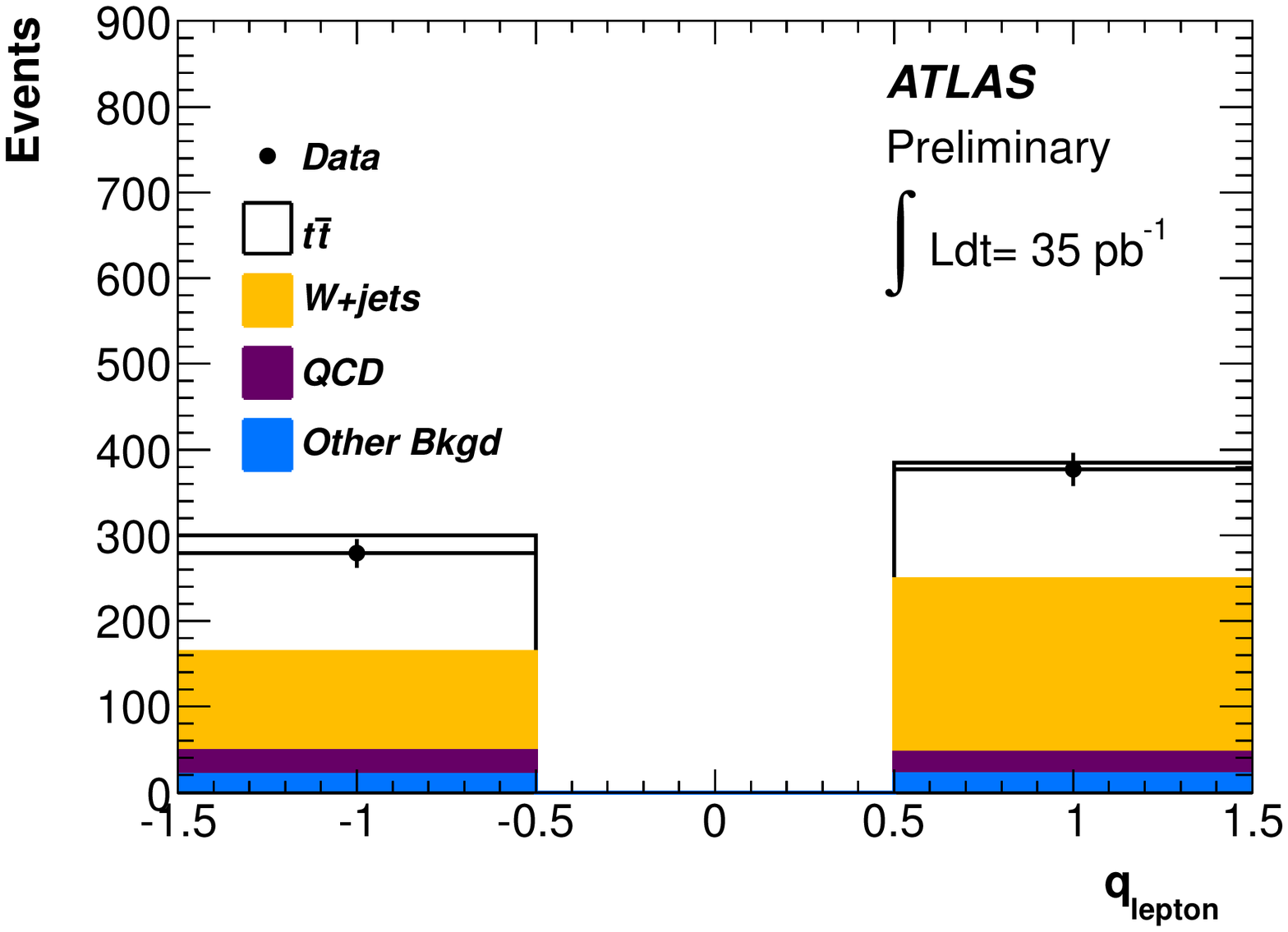}

(c)\\
\includegraphics[width=0.9\textwidth,keepaspectratio=true]{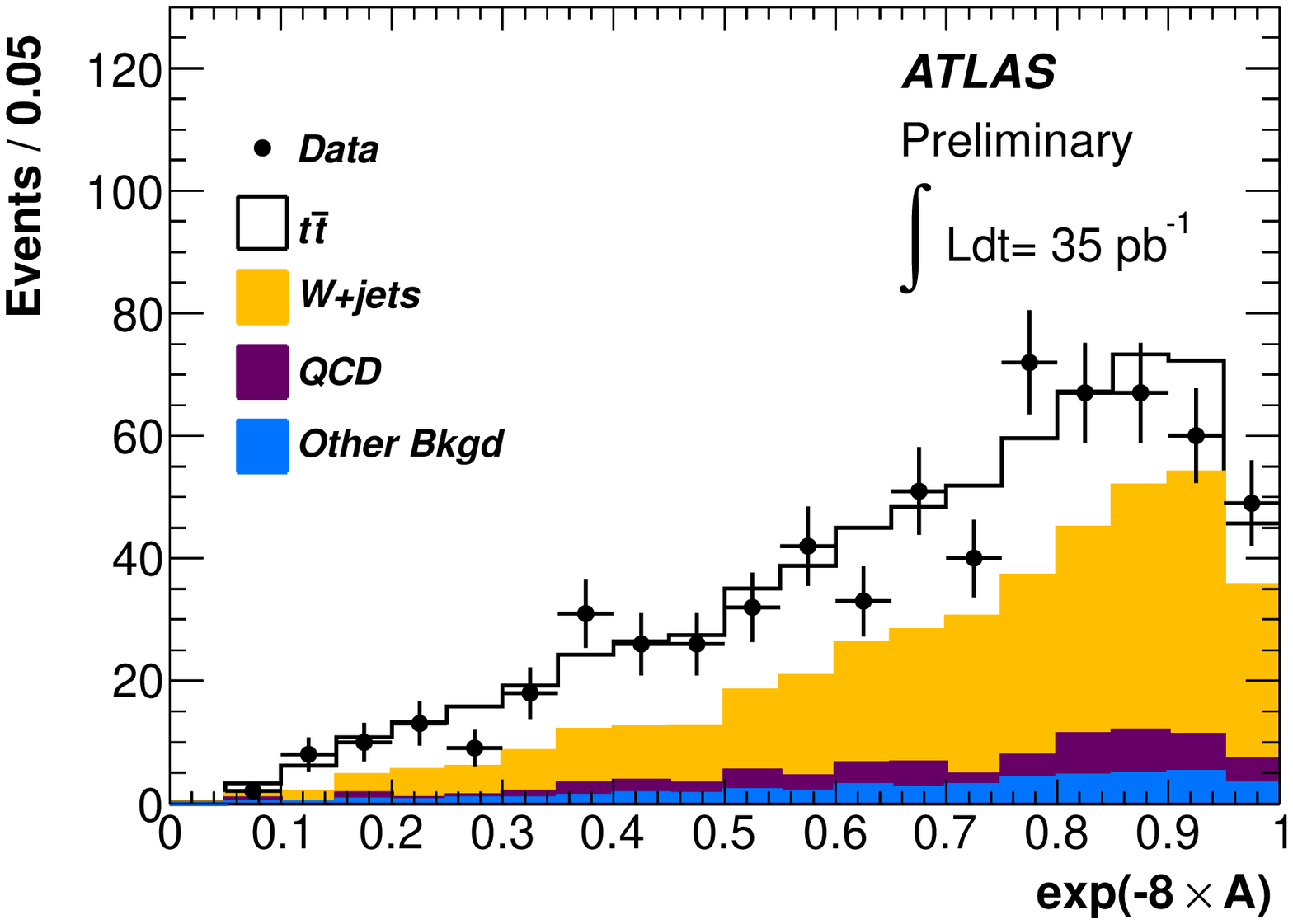}
\end{minipage}
\hfill
\begin{minipage}[t]{0.5\textwidth}
\centering
(b)\\
\includegraphics[width=0.9\textwidth,keepaspectratio=true]{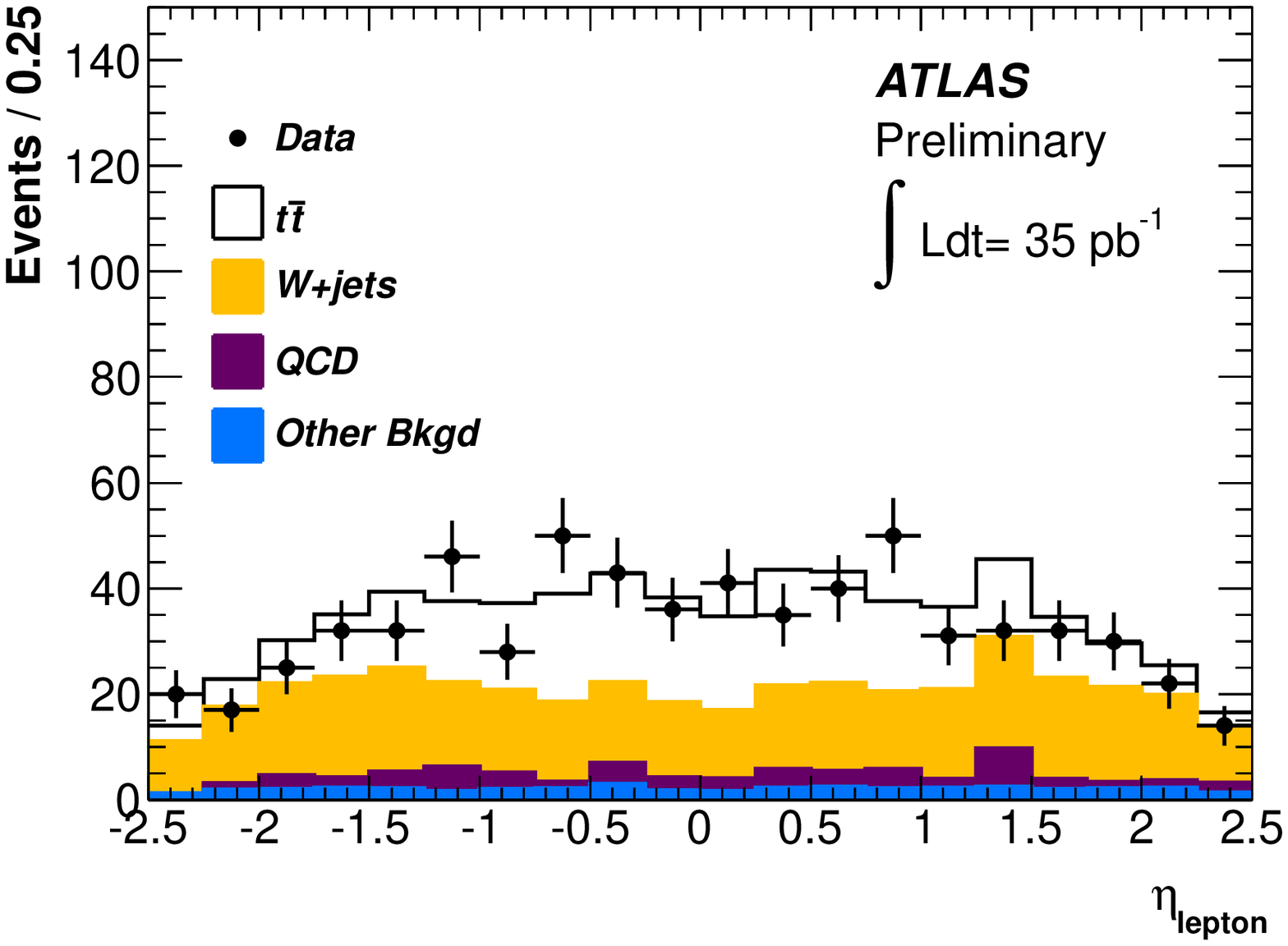}

(d)\\
\includegraphics[width=0.75\textwidth,keepaspectratio=true]{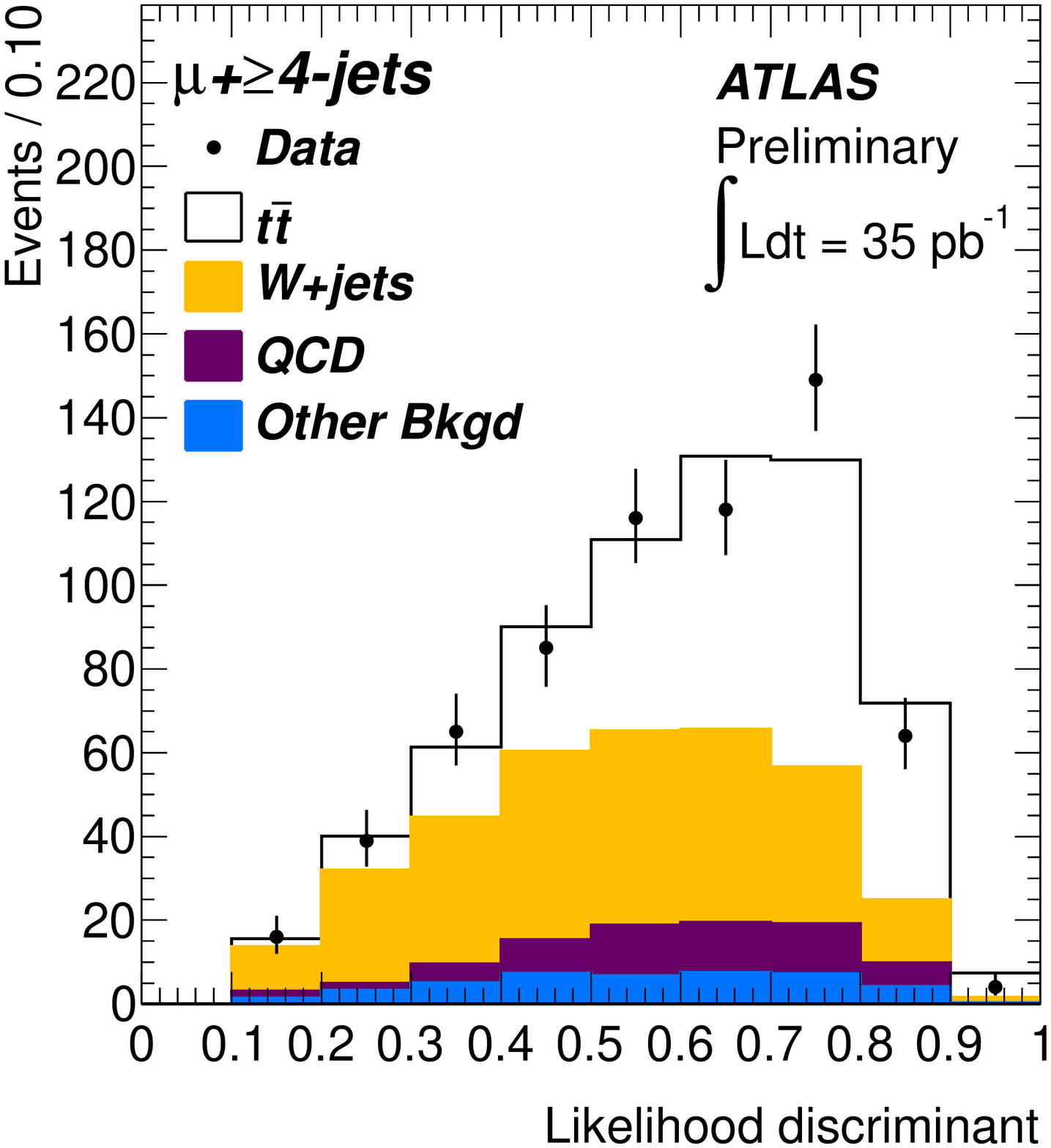} 
\end{minipage}
\caption{
Distributions of the input variables to the likelihood discriminant (a, b and c) 
and of the likelihood discriminant itself (d) 
for the $\mu$+$\geq$4-jets channel. 
In (a), (b) and (c) the normalizations for $t\bar{t}$ and $W$+jets are fixed to the theoretical predictions, 
while in (d) they are rescaled according to the result of the fit. 
The ``Other Bkgd'' (including $Z$+jets, single top and diboson) contribution is taken from theoretical predictions.
For QCD multi-jet the data-driven estimate is used.
}
 \label{fig:variables}
\end{figure}

A likelihood discriminant is built from these input variables following the projective likelihood approach
defined in the TMVA package \cite{tmva}. 
The distributions of the three input variables and of the likelihood discriminant
in data and simulated events are shown in Fig. \ref{fig:variables}, for the $\mu$+jets channel only.

A binned maximum likelihood fit is applied to the discriminant shapes to extract 
the $t\bar{t}$ cross-section. 
Likelihood functions are defined for each of the four channels 
($e$ and $\mu$, 3-jets and $\geq4$-jets) 
and are multiplied together in a combined fit to extract the total number of $t\bar{t}$ events. 

The performance of the likelihood fit (including statistical and systematic uncertainties) 
is estimated by performing pseudo-experiments. 
The systematic uncertainties associated with the simulation, object definitions and the QCD multi-jet estimate, 
as well as the statistical uncertainty and the uncertainty on the luminosity are  summarized 
in Table \ref{tab:syst}.

The result coming from the combined fit (including systematic uncertainties) is:
\begin{equation}
\sigma_{t\bar{t}} = 171 \pm 17 (stat.) ^{+20}_{-17} (syst.) \pm 6(lumi.) pb,
\end{equation}
for a total relative uncertainty of $-14.5/+15.5\%$.
The measured cross-section is in good agreement with the theoretical predictions.

\end{document}